\documentclass[oribibl]{llncs}

\usepackage{macros-ercim01}
\usepackage{graphicx}
\RequirePackage[ps2pdf,bookmarks=false,%
                urlcolor=blue,citecolor=blue,linkcolor=blue,%
                pagecolor=blue,%
                colorlinks,hyperfigures,pdfpagescrop={0 0 595 890}]{hyperref}

\title{Component Programming and Interoperability  \\
  in Constraint Solver Design}
\titlerunning{Component Programming and Interoperability
  in Constraint Solver Design}
\author{Fr\'ed\'eric Goualard%
  \thanks{The author acknowledges the financial support of the
    ERCIM Consortium in the form of a post-doctoral fellowship for the
    research exposed in this paper.}}
\institute{%
  Centrum voor Wiskunde en Informatica \\
  P.O.\ Box 94079,  NL-1090 GB AMSTERDAM\\
  \email{Frederic.Goualard@cwi.nl}}

\begin{document}
\maketitle

\begin{abstract}
  Prolog was once the main host for implementing constraint solvers.
  It seems that it is no longer so. To be useful, constraint solvers
  have to be integrable into industrial applications written in
  imperative or object-oriented languages; to be efficient, they have
  to interact with other solvers. To meet these requirements, many
  solvers are now implemented in the form of extensible
  object-oriented libraries. Following Pfister and Szyperski, we argue
  that ``objects are not enough,'' and we propose to design solvers as
  component-oriented libraries. We illustrate our approach by the
  description of the architecture of a prototype, and we assess its
  strong points and weaknesses.
\end{abstract}

\section{Introduction}
%%====================

From the 1980's onward, constraint programming techniques have been
considered a generalization of logic programming algorithms. The
use of Prolog as a hosting language for constraint solvers was then
natural. The principles of \emph{constraint logic programming} have been
embodied by many systems, such as Prolog~IV~\cite{Benhamou:25},
CHIP~\cite{Dincbas:01}, GNU Prolog (formerly
clp(FD))~\cite{Codognet:03} and \eclipse~\cite{Aggoun:03}, to name a
few.

At first sight, Prolog seems to be the ideal host for constraint
programming: it enforces declarative programming, which makes it a
neat modelling language, and it offers an elegant way of
handling combinatorial problems thanks to the availability of
nondeterminism at the language level. Prolog, however, suffers from
two flaws:
\begin{itemize}
\item it is far from being widely used in the software industry. As a
  consequence, the integration of constraint solving facilities in an
  application, most probably written in a different language, requires
  \textit{ad hoc} bridges and additional expertise in Prolog
  programming from the developers;
\item cooperation of different constraint solvers is the key to solve
  hard problems.  However, a Prolog interpreter/compiler is a
  closed system. Adding a new solving algorithm usually demands its
  direct implementation into the system, and the use of its internal
  data structures.  At best, Prolog-based environments
  are extensible. They do not offer \emph{interoperability},
  that is the possibility to communicate and use software components
  written independently.
\end{itemize}

In the beginning of the 1990's, Puget~\cite{Puget:03} tackled the
first point above by showing that it is possible to benefit from the
salient features of Prolog in a mainstream programming language,
namely \CPP. Since Puget's library Ilog Solver, many object-oriented
libraries have been devised for solving constraints:
IN\CPP~\cite{Hyvonen:08}, QDOOCS~\cite{Wiese:01},
OpAC~\cite{Goualard:10}\dots\ Most of them offer original features;
being libraries, all of them are easily extensible. Nevertheless, none
of them addresses the second point, that is interoperability. Every
library has its unique internal data structures, and consequently,
reusing directly solving methods from different libraries is not
possible. Yet, as said previously, the cooperation of solvers with
different strengths is the key to solve hard
problems~\cite{Benhamou:09,Monfroy-cosac,Marti:01}.  Reimplementing all
required methods in one system is too inefficient to be a viable
solution; in addition, which solver engine should be used?

The problem of making cooperate different solvers has long been
recognized and addressed in different ways. We present a critical
overview of some of the proposed solutions in
Section~\ref{sec:cooperation-solvers}, after having introduced some
basics on constraint solving in
Section~\ref{sec:constraint-programming}.

The lack of interoperability between libraries is not an issue limited
to constraint programming. There was a time when enthusiastic
foretellers were predicting that object-oriented programming would
create a component market; applications would then be built by
connecting arbitrary components in the same way as one uses
Lego blocks.  Sadly enough, Pfister and
Szyperski~\cite{Pfister:01} have shown that the concept of object is
insufficient to ensure the creation of true components.  However,
solutions do exist, which make \emph{component programming} a dream
(almost) come true. We present in
Section~\ref{sec:component-programming} the ideas underlying component
programming, and we point out the advantages offered by this
paradigm. In particular, we show what constraint programming can gain
from component programming. To support our claim, we present in
Section~\ref{sec:alix} the prototype of a \CPP\ library that uses
component programming techniques. Finally, we discuss in
Section~\ref{sec:perspectives} the pros and cons of our
approach and we consider the possible future of \emph{component
  constraint programming}.

\section{A Short Conspectus on Constraint Programming}
%%==============================================
\label{sec:constraint-programming}

Even an overview of constraint programming at large would deserve a
whole paper on its own. Hence, we will only consider in
this section the framework that is currently supported by the library to be
described in Section~\ref{sec:alix}.  The reader should nevertheless
keep in mind that the approach proposed in this paper is fully
general.

A \emph{Constraint Satisfaction Problem} (CSP) is defined as follows: given
a finite set of variables $\GSet{V}=\{v_1,\dots,v_n\}$, a Cartesian
product of domains $\Vector{D}=D_1\times\cdots\times D_n$ (with
$v_i\in D_i$ for any value of $i$ in $\{1,\dots,n\}$), and a set of relations
$\GSet{C}=\{c_1,\dots,c_m\}$ between the variables in \GSet{V}
(\emph{constraints}), we seek for all the possible assignments to the
variables that satisfy the conjunction $c_1\wedge\cdots\wedge c_m$.

Some popular methods for solving CSPs involve \emph{local consistency}
enforcement. The use of local consistency notions for solving
constraints can be traced back to the observation by Fikes~\cite{Fikes:01}
that, given a constraint $c(v_1,v_2)$ between two variables $v_1$ and
$v_2$ taking their values in the discrete domains $D_1$ and $D_2$, it
is possible to discard a value $a$ in $D_1$ whenever there is no value
$b$ in $D_2$ such that $c(a,b)$ holds. This idea is at the root of the
various algorithms to enforce local consistencies, such as the Waltz
algorithm~\cite{Waltz:01} or AC3~\cite{Mackworth:01} for computing 
\emph{arc consistency} and \emph{hyper-arc consistency}.

\begin{definition}[Hyper-arc consistency]
Let $c(v_1,\dots,v_n)$ be a constraint, $i\in\{1,\dots,n\}$ an integer, 
and $\Vector{D}=D_1\times\cdots\times D_n$ a Cartesian product of domains.
The constraint $c$ is said \emph{hyper-arc consistent w.r.t.\
    \Vector{D}} and $v_i$ if
\begin{equation}\label{eq:hyper-arc-consistency}
D_i=\KthProj{i}{\CRel{c}\cap\Vector{D}}
\end{equation}
where \CRel{c} is the relation associated to the constraint $c$, and 
$\KthProj{i}{\rho_1\times\cdots\times\rho_{i-1}\times\rho_i\times\rho_{i+1}
  \times\cdots\times\rho_n}=\rho_i$, for $\rho_k$ ($k\in\{1,\dots,n\}$)
unary relations.
The constraint $c$ is said hyper-arc consistent w.r.t.\ \Vector{D} if
Eq.~\eqref{eq:hyper-arc-consistency} is verified for all $i$ in
$\{1,\dots,n\}$. A constraint system $\GSet{C}=\{c_1,\dots,c_m\}$ is hyper-arc
consistent if each $c_i$ ($i\in \{1,\dots,n\}$) is hyper-arc consistent.
\end{definition}

Enforcing hyper-arc consistency over a constraint system is done by a
propagation algorithm such as the one given in Table~\ref{alg:nc3},
which applies a contracting operator $N$ on each constraint in turn
until reaching a fix-point. Provided the operators $N$ fulfill some basic
properties, algorithms like \algo{NC3} are confluent and
terminating~\cite{Gusgen:01}.

%--------------------------------------------------------------------ALGORITHM-
\begin{table}[htbp]
\caption{The \algo{NC3} propagation algorithm}
\smallskip
\begin{tt}
\begin{tabbing}
12\=123\=123\=123\=\kill
\algo{NC3}(\IN\ $\{c_1,\dots,c_m\}$; 
           \INOUT\ $\Vector{D}=D_1\times\cdots\times D_n$) \\
\BEGIN \\
\>    $\GSet{S} \Gets \{c_1,\dots,c_m\}$ \REM{Constraints added to the propagation set}\\
\>    \WHILE{\GSet{S}\neq\emptyset\ANDTHEN\Vector{D}\neq\emptyset} \DO \\
\>\>        $c \Gets \text{choose one $c_i$ in \GSet{S}}$\\
\>\>        $\Vector{D'} \Gets N_c(\Vector{D})$ \REM{Enforcing local consistency on $c$ and \Vector{D}} \\
\>\>        \IF{\Vector{D'} \neq \Vector{D}} \THEN\\
\pushtabs
12\=123\=123\=$\GSet{S} \Gets \GSet{S}\cup\{c_j\mid \exists v_k\in \Var{c_j} \wedge I'_k\neq I_k \}$12\=\kill
\>\>\>          $\GSet{S} \Gets \GSet{S}\cup
                                    \{c_j\mid \exists x_k\in 
                         \Var{c_j} \wedge D'_k\neq D_k \}$ \>\REM{\Var{c_j}: set of variables}\\ 
\>\>\>         $\Vector{D} \Gets \Vector{D'}$ \>\REM{occurring in $c_j$}\\
\poptabs
\>\>        \FI\\
\>\>        $\GSet{S}\Gets\GSet{S}\setminus\{c\}$ \\
\>     \WEND \\
\END
\end{tabbing}
\end{tt}
\label{alg:nc3}
\end{table}
%------------------------------------------------------------------------------

Hyper-arc consistency can be expensive to compute for discrete
variables since it induces holes in their domains; in addition, it is
uncomputable for continuous variables. Therefore, several weakening of
hyper-arc consistency have been defined: \emph{bounds
  consistency}~\cite{Marriott:01} where only the bounds of the domains are
considered, \emph{box consistency}~\cite{Benhamou:02} for
continuous CSPs\dots\ Due to lack of space, we will not present these
consistencies and the reader is referred to the references given for
additional information.

\section{Making Solvers Cooperate}
%%================================
\label{sec:cooperation-solvers}

In the following, we use the term \emph{solver} in a broad sense,
namely: a solver is any ``box'' that takes as input some constraints
and the domains of the variables involved, and that returns as output
a set of constraints that defines the same solution set as the input
ones and the domains of the variables that have been possibly
tightened. Hence, we consider as solvers procedures that compute
redundant constraints, or that simplify them. Most solvers, however,
will leave the constraint set unchanged and will simply narrow down
the domains of the variables.

Solvers can be limited to some particular kind of constraints (the
simplex method for linear equalities, for instance); alternatively,
some costly solvers are best used in conjunction with other simpler
ones (symbolic methods and interval methods, for
instance~\cite{Benhamou:09}). The cooperation of solvers has been
shown to be a key concept in solving hard problems. One way to take
this fact into account is to tightly integrate several solvers in one
system (see Prolog~IV~\cite{Benhamou:25}). However, such a
centralization hinders the development of new methods. Another way is
to have different independent tools cooperate. 

Several schemes have been devised for allowing such a cooperation
between systems with different inner structures.
\cosac~\cite{Monfroy-cosac}, for instance, is a client/server
architecture to manage the exchange of information among different
solvers through \emph{pipes}. The data exchanged are character
strings. In the same spirit, AMPL~\cite{Gay:02} allows one to hook a
new solver to the system through a very simple interface. The AMPL
kernel supervises the exchange of information through files in a
particular format. Both systems are interesting in that they are
truly open; they allow cooperation between completely independent
solvers. The amount of work to add a new solver only
requires the addition of a small interface to have it understand the
format used for inputs and outputs. However, the cooperation happens
at a fixed level and a solver cannot obtain additional information
from the internal structure of the constraint store that would allow
it to speed up the solving process. In addition, the use of a
primitive format for information interchange, namely text, incurs a
significant slow down. As a consequence, communications must be kept
to a minimum. A side effect is also that nothing can prevent a priori
the connection of a solver that do not understand correctly the input
format (no type safety). Lastly, these schemes do not allow distributed 
cooperation.

To overcome some of these problems, Ng \textit{et al.}~\cite{NgKB:01}
have proposed a generic \CPP\ interface for connecting systems such as
Ilog Solver~\cite{Puget:03} and Oz~\cite{Smolka:02}.  
The communication takes place at the lowest
level, which ensures performances, and type safety is guaranteed by
the language. A major inconvenience lies in that this communication
scheme often requires the extension of the systems to provide some
necessary hooks. On the technical level, all solvers have to be
written in \CPP\ and compiled with the same compiler to be able to
communicate. In addition, the level of cooperation is once again fixed 
and depends on the interface specifications.

In fact, the lack of interoperability is a language issue that is not
limited to constraint programming systems. For many years, it was
believed that object-oriented programming would permit the development
of interoperable components. Pfister and Szyperski~\cite{Pfister:01}
have shown that it is not the case. However, a new paradigm,
\emph{component programming}, has emerged, that allows us to meet this
goal.

\section{Component Programming}
%%=============================
\label{sec:component-programming}

In an object-oriented language, communication is done by method calls
between objects. Usually, an object cannot be used in isolation
because it does not represent a concept and it relies on services
provided by other objects. In addition, objects have to know which
objects they must call to send and receive information.

By contrast, Szyperski~\cite{Szyperski:02} defines components as
\emph{units of deployment}. They may be composed of one or several
objects; they might even be some non-object-oriented piece of code.

The communication model among components is based on \emph{slots} and
\emph{signals}. A component whose state has been modified sends the
information as a \emph{signal} (an \emph{event} in the Java
terminology) to the other components that might be interested in it. A
component that wants to be informed of the modification of another
component (a \emph{listener} in Java terminology) connects its input
slot to the output slot of that component.  It is important to stress
that the sender does not know anything about the components listening
to its signals. It even does not know whether anybody is listening at
all. On the other side, the receiver does not have to know the precise
type of the sender.  Suffice that the sender provides the right kind
of signal. Several components can listen to the same slot, and a
component can listen to several slots. In addition, a component can be
dynamically connected to, and disconnected from a slot. Lastly,
two connected components can indifferently run on the same or on 
different computers; this is transparent to the communication process.

The communication model by signals and slots is already extensively
used in Graphical User Interface libraries, such as Qt, the GUI layer
for the KDE environment (\href{www.kde.org}{www.kde.org}), awt and Beans in
Java, or Gtk, the layer for the Gnome environment
(\href{www.gnome.org}{www.gnome.org}). It is also part of
Microsoft's \emph{Component Object Model}~\cite{Williams:01}.

There are few languages that offer a native support for component
programming. However, component programming can be achieved at a
reasonable cost with most object-oriented languages such as \CPP\ and
Java. 

Components can be freely connected provided they agree on the kind of
messages they want to exchange (type safety). They can communicate
transparently in a distributed environment. The cost of
communications is reasonably cheap since the communication mecanism is
at the same level as the rest of the code. Lastly, protocols such as
CORBA~\cite{OMG:01} allow communications between programs written in
different languages.

All these qualities led us to devise aLiX (\emph{a Library for
  Constraint Solving}), a \CPP\ constraint solving library based on
the component programming paradigm. The description of its
architecture is the subject of the next section; The impact of this
approach on constraint solver design and the new possibilities offered
are discussed in Section~\ref{sec:perspectives}.

\section{The aLiX architecture}
%%=============================
\label{sec:alix}

The implementation of aLiX is still a work in progress, though we
already have a full system that allowed us to validate our approach.
Due to lack of space, we only skim over the salient features of the
aLiX architecture to show the new possibilities offered by the
component programming paradigm.

\begin{figure}[htbp]
  \begin{center}
    \includegraphics[width=.8\textwidth]{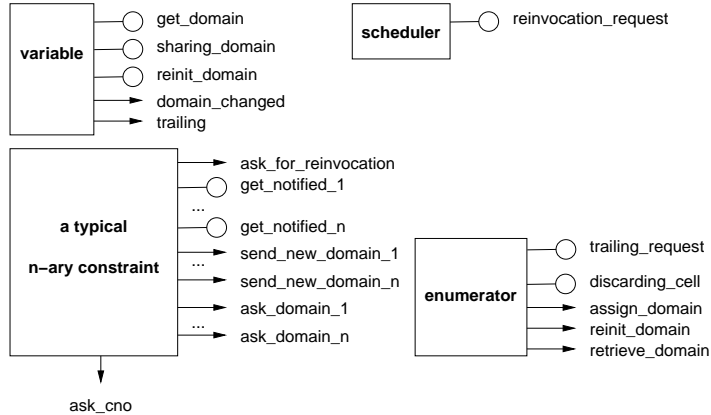}
    \caption{Core of aLiX}
    \label{fig:core-alix}
  \end{center}
\end{figure}

The core of aLiX is composed of four main concepts (see
Fig.~\ref{fig:core-alix}):
\begin{description}
\item[variables.] The base class for the variables offers three input
  slots and two output slots:
  \begin{description}
  \item[\ttfamily get\_domain.] This is the access point to modify the domain 
    of the variable,
  \item[\ttfamily sharing\_domain.] This is an input slot used for
    sending information outside (cf. the comment by
    Szyperski~\cite[p.\ 149]{Szyperski:02}). Components send a message to this
    slot which is used to encapsulate and retrieve the current domain of the
    variable,
  \item[\ttfamily reinit\_domain.] The variable reinits its domain with
    the one sent to this slot. The difference with the
    \code{get\_domain} access point lies in that such an assignment
    is usually not considered as an event (though this can be modified by
    inheriting from the base class and redefining the handler of the 
    slot),
  \item[\ttfamily trailing.] Each time the variable is about to modify
    its domain, it first sends its current domain through this slot. 
    A backtrack stack is usually connected to this slot just before 
    enumerating the values of the variable's domain,
  \item[\ttfamily domain\_changed.] After any modification, the variable 
    sends its new domain through this slot;
  \end{description}
  One can inherit from the \code{variable} base class to add new
  slots. For example, aLiX contains the class
  \code{integral\_variable} (variable with a discrete domain), which
  uses three additional output slots to send messages whenever the
  hull of the domain, the left bound, or the right bound have been
  changed, and one slot to send a message upon instantiation of the
  variable.
\item[constraints.] The class associated to this concept is abstract
  and is only used to offer some facilities such as a unique identifier.
  Each constraint to be solved is represented by
  a constraint object that connects itself to the
  \code{domain\_changed} (or \code{hull\_changed},\dots\ for that
  matter) slot of all the variables involved in the relation
  (\code{get\_notified} slots). It possesses also some output slots to
  be connected to \emph{narrowing objects} (cnos) to create two-ways
  channels.  A narrowing object has input slots to receive domains. It
  reduces these domains by enforcing some consistency and returns them
  through the channel. One constraint may use several narrowing
  operators at a time (with different tightening abilities); it can
  also disconnect itself from some cnos and connect itself to other
  cnos dynamically to choose the best suited cnos at any moment. The
  message exchanged with the cnos contains the domains to be tightened
  and a flag to be raised by a cno whenever some failure occurs. The
  output slot is \emph{marshalled} in such a way that the dispatching
  of the domains to the other cnos listening to the slot is
  immediately stopped in case of failure. Whenever a constraint
  receives a message from a variable whose domain has been modified,
  it has the possibility to send some messages to the cnos to reduce
  immediately the domains of the other variables; alternatively, it
  can send a message through its \code{ask\_for\_reinvocation} slot to
  be managed by a scheduler (see below);
\item[schedulers.] A scheduler is an object centralizing the requests
  for reinvocation of the constraints. It is used to implement
  propagation algorithms such as AC3~\cite{Mackworth:01}. However, a
  scheduler is fairly general in that it can handle indifferently
  constraints, variables, or any other object inheriting from the
  \code{schedulable} class. Such a flexibility allows us to use a
  constraint-oriented propagation scheme (where the propagation queue
  contains constraints) or a variable-oriented propagation scheme
  (where the propagation queue contains variable whose domain has been
  modified) as implemented in clp(FD)~\cite{Codognet:03} by simply
  selecting different connection schemata;
\item[enumerators.] Once all the constraints have been added to the
  store, an enumerator is used to separate the different solutions and
  to overcome the incompleteness of the consistencies enforced. A
  discrete enumerator assigns to each variable every value in its
  domain and reinvokes the propagation process to check the
  consistency of the assignment. A continuous enumerator splits
  recursively the domains of the variables (cf.\ the \code{solve}
  procedure available in many CLP systems). An enumerator is also
  responsible for managing a backtrack stack on which are saved the
  domains of the variables to be recovered upon backtracking. The user
  can define different enumerators to select a particular heuristics
  for choosing the next variable to be considered, for instance.
\end{description}

%%-------------------------------------------------------------------- Figure -
%%-----------------------------------------------------------------------------

At present, aLiX relies on the ADDL library~\cite{Goualard:12}
(\emph{A Discrete Domain Library}) for representing the domains of the
discrete variables. This library offers both connected domains
(intervals) and domains with holes. Domains are considered as STL
containers and offer STL-like iterators to retrieve values, which
permits to consider connected and disconnected domains in a uniform
way. Disconnected domains use a reference-counting mecanism with
copy-on-write semantics (see the C++ FAQ Lite at
\href{http://www.cerfnet.com/~mpcline/c++-faq-lite/}{http://www.cerfnet.com/\~{}mpcline/c++-faq-lite/})
to ensure the best performances.

Table~\ref{tab:n-queens} presents an aLiX program to solve the $n$ queens
problems. This program uses disconnected domains and a variable-oriented
propagation scheme as in GNU Prolog. It implements the naive algorithm
that is a direct representation of the specifications of the problem. We have
also tested some other formulations proposed by Puget that use global
\code{alldiff} constraints and a constraint-oriented propagation scheme. We
have then be able to check that the coding of the alternative solutions
only required a different connection schema between the constraints and the
variables.

%%-------------------------------------------------------------------- Table -
\begin{table}[htbp]
\begin{multicols}{2}\tiny
%
% This file was automatically produced at Oct  3 2000, 20:14:39 by
% c++2latex queens.cpp
%
\expandafter\ifx\csname indentation\endcsname\relax%
\newlength{\indentation}\fi
\setlength{\indentation}{0.5em}
\begin{flushleft}
{$//$\it{} Constraint narrowing operator for the variable x{}\mbox{}\\
}{\bf class} out\_of : {\bf public} cno \{\mbox{}\\
{\bf public}:\mbox{}\\
\hspace*{2\indentation}out\_of(fd\_variable\& x, fd\_variable\& y, {\bf int} c) \{\mbox{}\\
\hspace*{4\indentation}connect(x.get\_domain,send\_domain);\mbox{}\\
\hspace*{4\indentation}connect(x.sharing\_domain,ask\_domain\_x);\mbox{}\\
\hspace*{4\indentation}connect(y.sharing\_domain,ask\_domain\_y);\mbox{}\\
\hspace*{4\indentation}cst=c;\mbox{}\\
\hspace*{2\indentation}\}\mbox{}\\
\hspace*{2\indentation}bool invoke() \{ {$//$\it{} Called when y is instantiated{}\mbox{}\\
}\hspace*{4\indentation}{$//$\it{} Retrieving current domain of x{}\mbox{}\\
}\hspace*{4\indentation}ask\_domain\_x.share(dom\_x); \mbox{}\\
\hspace*{4\indentation}ask\_domain\_y.share(dom\_y); \mbox{}\\
\hspace*{4\indentation}{$//$\it{} x $<$- x$\backslash$\{d,d$+$cst,d-cst\}{}\mbox{}\\
}\hspace*{4\indentation}((dom\_x \%= d) \%= d+cst) \%= d-cst;\mbox{}\\
\hspace*{4\indentation}send\_domain.send(dom\_x); {$//$\it{} Sending new domain to x{}\mbox{}\\
}\hspace*{4\indentation}{\bf return} dom\_x.is\_not\_empty();\mbox{}\\
\hspace*{2\indentation}\}\mbox{}\\
output\_slots:\mbox{}\\
\hspace*{2\indentation}output\_slot$<$fd\_domain$>$ send\_domain;\mbox{}\\
\hspace*{2\indentation}output\_share\_slot$<$fd\_domain$>$ ask\_domain\_x, ask\_domain\_y;\mbox{}\\
{\bf private}:\mbox{}\\
\hspace*{2\indentation}fd\_domain doma\_x, dom\_y;\mbox{}\\
\hspace*{2\indentation}{\bf int} cst;\mbox{}\\
\};\mbox{}\\
\mbox{}\\
{$//$\it{} \{x$<$$>$y, x$<$$>$y$+$i, x$<$$>$y-i\} in one constraint{}\mbox{}\\
}{\bf class} diff3 : {\bf public} constraint \{\mbox{}\\
{\bf public}:\mbox{}\\
\hspace*{2\indentation}diff3(fd\_variable\& x, fd\_variable\& y, {\bf int} i) \{\mbox{}\\
\hspace*{4\indentation}scheduled.insert(scheduled.begin(),\&x);\mbox{}\\
\hspace*{4\indentation}scheduled.insert(scheduled.begin(),\&y);\mbox{}\\
\hspace*{4\indentation}x.on\_instantiation({\bf new} out\_of(y,x,i));\mbox{}\\
\hspace*{4\indentation}y.on\_instantiation({\bf new} out\_of(x,y,i));\mbox{}\\
\hspace*{2\indentation}\}\mbox{}\\
\};\mbox{}\\
\mbox{}\\
{\bf int} main() \{\mbox{}\\
\hspace*{2\indentation}{\bf int} n; cout $\ll$ {\tt"N? "}; cin $\gg$ n;\mbox{}\\
\mbox{}\\
\hspace*{2\indentation}Vector$<$fd\_variable$\ast$$>$ x(1,n);\mbox{}\\
\hspace*{2\indentation}fifo\_scheduler store;\mbox{}\\
\mbox{}\\
\hspace*{2\indentation}{\bf for} ({\bf int} i=1;i$<$=n;++i) \{\mbox{}\\
\hspace*{4\indentation}x.insert(x.begin(),\mbox{}\\
\hspace*{13\indentation}{\bf new} fd\_variable(fd\_domain(1,n)));\mbox{}\\
\hspace*{2\indentation}\}\mbox{}\\
\hspace*{2\indentation}enumerator\_round\_robin$<$fd\_domain$>$ \mbox{}\\
\hspace*{4\indentation}inst(store,x.begin(),x.end());\mbox{}\\
\mbox{}\\
\hspace*{2\indentation}{\bf for} ({\bf int} i=1;i$<$=n;++i) \{\mbox{}\\
\hspace*{4\indentation}{\bf int} k=1;\mbox{}\\
\hspace*{4\indentation}{\bf for} ({\bf int} j=i+1;j$<$=n;++j) \{\mbox{}\\
\hspace*{6\indentation}store.post($\ast$({\bf new} diff3($\ast$(x[i]),$\ast$(x[j]),k)));\mbox{}\\
\hspace*{6\indentation}++k;\mbox{}\\
\hspace*{4\indentation}\}\mbox{}\\
\hspace*{2\indentation}\}\mbox{}\\
\hspace*{2\indentation}{\bf if} (store.run()) \{\mbox{}\\
\hspace*{4\indentation}{\bf if} (inst.first\_solution()) \{\mbox{}\\
\hspace*{6\indentation}{$//$\it{} [Display domains]{}\mbox{}\\
}\hspace*{6\indentation}{\bf while} (inst.next\_solution()) \{\mbox{}\\
\hspace*{8\indentation}{$//$\it{} [Display domains]{}\mbox{}\\
}\hspace*{6\indentation}\}\mbox{}\\
\hspace*{6\indentation}cout $\ll$ {\tt"No more solutions"} $\ll$ endl;\mbox{}\\
\hspace*{4\indentation}\} {\bf else} \{\mbox{}\\
\hspace*{6\indentation}cout $\ll$ {\tt"No solution"} $\ll$ endl;\mbox{}\\
\hspace*{4\indentation}\}\mbox{}\\
\hspace*{2\indentation}\}\mbox{}\\
\}\mbox{}\\
\end{flushleft}

\end{multicols}
\caption{The $n$ queens problem formulated in aLiX}
\label{tab:n-queens}
\end{table}

\section{Discussion and Perspectives}
%%===================================
\label{sec:perspectives}

We have tested the current version of aLiX on several standard
benchmarks such as the $n$ queens problem. It appears that our library
is roughly three times slower than GNU Prolog 1.2.1 and three times
faster than \eclipse\ 5.1.0 on the same formulations of the problems.
We have however identified several points where optimization might
dramatically speed-up the computation. All the tests have been
performed on discrete CSPs. Addition of components for solving
continuous CSPs is one of our priority for the near future. We believe
that the overhead incurred by the more elaborate communication scheme
will be less acute for this kind of problems since the time spent in
the narrowing objects is likely to overwhelm the one spent in
communications.

Though still at an early stage, we have been able to check the
versatility of our approach by writing different formulations for the
same benchmarks and changing the propagation scheme by mere
reorganization of the connections between the objects. 

Slots offer natural hooks for spying any event occurring inside a
solver engine. There is a large number of potential uses of this for writing
non-intrusive debuggers or animations of the solving of some problems
for educational purpose, for instance. Suffice for the applications to
connect themselves to the appropriate slots without interfering in any way
with the solver.

The current version of aLiX uses ad hoc classes for implementing the
signal/slot mechanism. We are in the process of replacing the use of
these classes by the library sigc++
(\href{libsigc.sourceforge.net}{libsigc.sourceforge.net}). This library 
is the one used in Gtk for connecting components. An extension of it
already exists to allow communications in a distributed environment. Another
extension is planned in the near future to support the CORBA protocol.
By using sigc++, aLiX will offer both the possibility to solve constraints
in a distributed environment in a transparent way, and the ability to use
programs implementing narrowing operators written in a language different
from \CPP.

The use of an \emph{Object Request Broker} greatly augments
interoperability but it incurs a non-negligeable cost. As a
consequence, we believe that its use should be only considered when
the time spent in actual computation in the components is large enough
to make the communication time insignificant. Such a situation is most
likely to arise when components implement narrowing operators for
global constraints or costly processes such as factorization of the
expression of the constraints, or computation of a Gr\"obner basis, for
instance.

The interoperability offered by the component programming approach can
be used at various levels. However, the most visible one seems to lie
in the possible plugging of third party narrowing objects (be they
real objects or complete applications) onto constraints. At present,
narrowing objects have to use the same arithmetic library as the one
used by constraints since messages between them are domains. It is however
possible to allow a more stringent separation by exchanging
messages containing \emph{OpenMath code}
(\href{www.openmath.org}{www.openmath.org}) or MP packets
(\href{http://www.symbolicnet.org/areas/protocols/mp.html}{http://www.symbolicnet.org/areas/protocols/mp.html}),
for instance.

\section*{Acknowledgements}
%%========================
The author thanks Laurent Granvilliers for interesting discussions on
the topic of event-driven constraint solver design, Maarten Van Emden
for discussions on component programming, and \'Eric Monfroy for
pointing out overlooked references in the area of cooperative
constraint solving and for sharing his insight on distributed
constraint solving.

%% Bibliography
%%=============
\bibliographystyle{plain}
\bibliography{goualard-ercim01}

\end{document}